\documentclass{sig-alternate}
\usepackage{times}

\usepackage[utf8]{inputenc}
\usepackage[T1]{fontenc}

\usepackage[caption=false,font=footnotesize]{subfig}

\usepackage{floatrow}
\usepackage[table, dvipsnames]{xcolor}  
\usepackage{fancyhdr}
\usepackage{rotating}
\usepackage{graphicx}
\usepackage{url}
\usepackage{cite}
\usepackage{verbatim}
\usepackage{color}
\usepackage{colortbl}
\usepackage{multirow}
\usepackage{hhline}
\usepackage{amssymb}
\usepackage[us]{datetime}
\usepackage{amsmath}        

\usepackage{paralist}
\usepackage{stfloats}
\usepackage[ruled, vlined, norelsize]{algorithm2e} 
\SetAlFnt{\sf} 

\graphicspath{{incl/}}

\newdimen\arrayruleHwidth
\makeatletter
\def\Hline{\noalign{\ifnum0=`}\fi\hrule \@height \arrayruleHwidth
\futurelet \@tempa\@xhline}
\makeatother

\makeatletter
\def\blfootnote{\xdef\@thefnmark{}\@footnotetext}
\makeatother

\usepackage{setspace}
\def\smallerspacecaption{\vspace{-5mm}}

\definecolor{gray}{gray}{0.9}

\newcommand{\drop}[1]{}

\shortdate
\settimeformat{ampmtime}

\makeatletter
\newcommand*\refsize{%
	   \@setfontsize\refsize{7}{7.5}%
}
\makeatother

\def\sharedaffiliation{%
\end{tabular}
\renewcommand{\arraystretch}{1.4}
\begin{tabular}{c}}

\begin{document}

\setlength{\textfloatsep}{9pt plus 2pt minus 4pt}
\setlength{\floatsep}{6pt plus 2pt minus 2pt}
\setlength{\dbltextfloatsep}{9pt plus 2pt minus 4pt}
\setlength{\dblfloatsep}{6pt plus 2pt minus 2pt}

\CopyrightYear{2017}
\setcopyright{acmcopyright}
\conferenceinfo{DAC '17,}{June 18-22, 2017, Austin, TX, USA}
\isbn{978-1-4503-4927-7/17/06}\acmPrice{\$15.00}
\doi{http://dx.doi.org/10.1145/3061639.3062293}

\title{
	On Mitigation of Side-Channel Attacks in 3D ICs:
		Decorrelating Thermal Patterns from Power and Activity
}

%
\numberofauthors{2} 
\author{
\alignauthor
Johann Knechtel\\
\alignauthor
Ozgur Sinanoglu\titlenote{The second author is partially supported by the Center for Cyber Security (CCS) at NYUAD.}\\
\sharedaffiliation
\email{\{johann, ozgursin\}@nyu.edu}\\
\affaddr{Electrical and Computer Engineering, New York University Abu Dhabi (NYUAD), United Arab Emirates}
}

\maketitle

\renewcommand{\headrulewidth}{0.0pt}
\thispagestyle{fancy}
\chead{
\copyright~ACM, 2017. This is the author's version of the work. It is posted here by permission of
	ACM for your personal use. Not for redistribution. The definitive version was
	published in Proc. Design Automation Conference, 2017
http://dx.doi.org/10.1145/3061639.3062293
}

\begin{abstract}
Various side-channel attacks (SCAs) on ICs have been successfully demonstrated and also mitigated to some degree.  In the context of 3D ICs, however, prior art has mainly
focused on efficient implementations of classical SCA countermeasures.
That is, SCAs tailored for up-and-coming 3D ICs have been overlooked so far.
In this paper, we conduct such a novel study and focus on one of the most accessible and critical side channels: thermal leakage of activity and power patterns.
We address the thermal leakage in 3D ICs early on during floorplanning,
   along with tailored extensions for power and thermal management.
Our key idea
is
to carefully exploit the specifics of material and structural properties in 3D ICs, thereby decorrelating the thermal behaviour from underlying power and activity patterns.
Most importantly, we discuss powerful SCAs and demonstrate how our open-source tool helps to mitigate them.

\end{abstract}

%




\section{Introduction}
\label{sec:introduction}

{\bf Side channels} represent the physical interactions any electronic device triggers in its surrounding, e.g., electromagnetic radiation or heat dissipation.
{\bf Side-channel attacks (SCAs)} then seek to retrieve sensitive information (e.g., secret keys) by carefully monitoring those physical interactions~\cite{rostami14, zhou05,
	skorobogatov12}.
For example, Skorobogatov and Woods~\cite{skorobogatov12} extracted an embedded AES key from a military-grade FPGA
using an advanced attack on the power side channel.
Another
practical and effective
side channel
is the thermal
leakage of
power and
activity patterns~\cite{hutter14, masti15}.
Being the scope of our work,
this side channel is
elaborated in detail in Section~\ref{sec:scope}.

{\bf 3D integration} is
expected
to successfully meet the increasingly demanding requirements for modern ICs, such as 
high performance, increased functionality, and low power consumption~\cite{ITRS15, kim12_3dmaps, iyer16}.
While previous studies, early prototypes, and first commercial products have focused on memory-logic
integration (e.g., see~\cite{kim12_3dmaps, iyer16, beneventi16}), there is recent progress towards logic stacking as well, also
	    driven by
	    technology innovations such as
direct wafer bonding~\cite{ITRS15}.

One major concern for 3D integration in general and for 3D logic ICs in particular
is the thermal and power management, which is furthermore highly technology dependent~\cite{knechtel16_Challenges_ISPD, budhathoki16, samal16}.
In practice, 3D ICs will require runtime capabilities for thermal management, based on
	embedded on-chip thermal sensors~\cite{beneventi16,zhu08,fu17}.

{\bf SCAs in the context of 3D ICs} have only recently come into focus, and
studies such as~\cite{xie16} mainly discuss how structural properties of 3D
ICs
may help in mitigating SCAs. The full potential of SCAs specifically tailored for and targeting on 3D ICs has not been explored yet.
It is intuitive that SCAs developed for 2D ICs may not be successful in 3D ICs,
due to the different physical structures of 2D and 3D ICs.
Still, a practical SCA
	could be based on thermal readings of 3D ICs, which is realistic
	due to the availability of and relatively easy access to
	on-chip thermal sensors~\cite{masti15,beneventi16,zhu08,fu17}.

{\bf
SCA countermeasures in 3D ICs}
such as~\cite{bao15, valamehr10} leverage 3D integration mainly for efficient implementation
of prior techniques.
For example, to mitigate timing-based SCAs on caches,
Bao and Srivastava~\cite{bao15} exploit the reduced latencies in 3D ICs in order to implement
	random eviction and latency scattering, which are otherwise (in the context of 2D ICs) arguably too costly measures.

The work of Gu {\em et al.}~\cite{gu16_ICCD} seeks to mitigate thermal-related leakage in 3D ICs, and is of particular interest here.
The authors propose to integrate (along with the actual functional circuitry) thermal sensors, thermal-noise generators as well as runtime controllers.
These primitives
``inject dummy activities'' when-/wherever considered necessary and, hence, aim for smooth thermal profiles to hinder thermal profiling
of module activities.
However, there are major shortcomings of this work. First, the concept appears not specifically tailored for 3D ICs---it does not explicitly address the different structural
and material properties, different heat paths or thermal coupling effects in 3D ICs, nor does it elaborate
on the
implementation of 3D ICs.
Second, the ``injection'' principle
causes further power dissipation, which may be prohibitive for thermal- and power-constrained 3D ICs in the first place. This shortcoming is confirmed by an observation
in~\cite{gu16_ICCD} itself: the best leakage-mitigation rates are only achievable for the highest injection rates.

{\bf In this work}, we take a fundamentally different approach by considering security as a design parameter. We propose an SCA-aware floorplanning methodology that decorrelates the
thermal behaviour from power and activity patterns, protecting against a very practical and potentially effective threat for the security of 3D ICs.

\section{Background and Scope}
\label{sec:scope}

\subsection{Thermal Side Channel}
\label{sec:background_TSC}

The thermal side channel (TSC) has recently gained more attention, also thanks to case studies such as~\cite{hutter14, masti15}.
For example, Masti {\em et al.}~\cite{masti15} have shown for Intel Xeon multi-core processors
that ($i$)~process executions on one core can be detected in adjacent cores, and that ($ii$)~different processes, when scheduled by turns in one core, can build a
covert channel with up to 12.5 bit/s.

The TSC is
particularly attractive for three reasons: ($i$) it is easy to access via widely available on-chip sensors~\cite{masti15,beneventi16,zhu08,fu17};
($ii$) it provides internal and external leakage of activity/computation patterns
through thermal variations~\cite{hutter14,masti15}; and ($iii$) it may serve as proxy for the power side-channel using temperature-to-power interpolation techniques such as~\cite{
	paek13}.
     An attacker can obtain
localized thermal readings either directly or estimate them, e.g., using interpolation techniques.
We note that interpolation techniques have been demonstrated to achieve high-accuracy and high-resolution estimates of both power and
		temperatures in large-scale 3D
ICs~\cite{beneventi16}.

In general, an attacker can monitor the thermal patterns of an IC either
at regular runtime or when he/she applies specific input patterns, crafted to reveal the activity of sensitive components. We address both scenarios for the
attacker's capabilities and the SCAs we
formulate (Section~\ref{sec:attacks}) and evaluate (Section~\ref{sec:experiments}).

\begin{figure}[t]
\centering
	\includegraphics[width=.7\columnwidth]{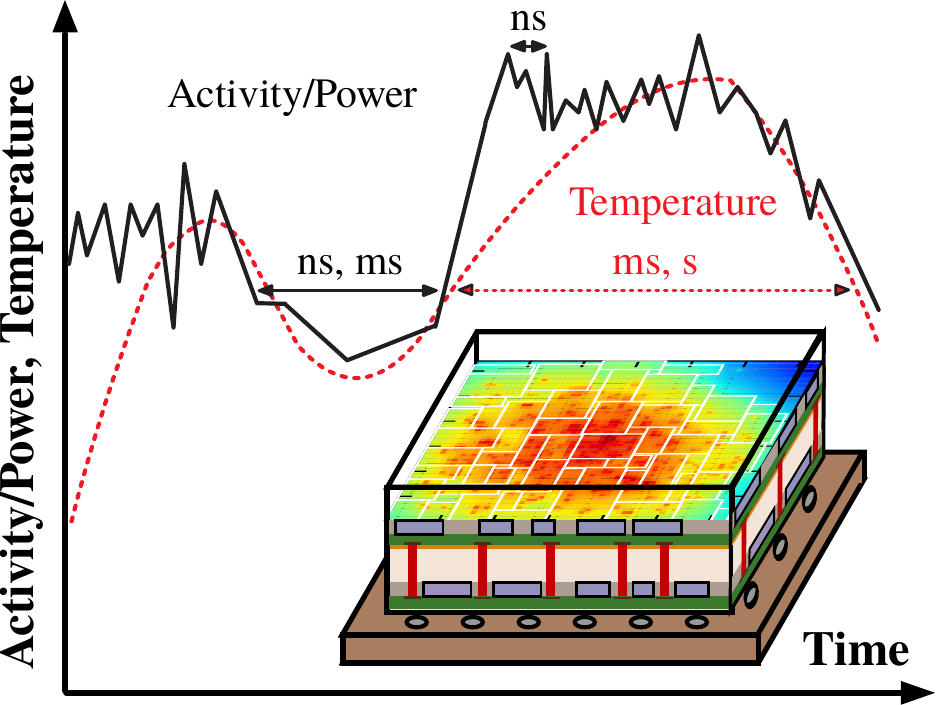}
	\smallerspacecaption
	\caption{The different time scales of activity/power (solid black curve) and temperature (dashed red curve) in ICs.
		In this paper, we focus on 3D ICs with two dies, stacked {\em face-to-back}.
    \label{fig:power-thermal-scales}
	}
\end{figure}

{\bf Practical limitations of the TSC} are the following:
\vspace{1.25mm}
\begin{compactitem}
\item The classical heat equation models power
	and temperature as linearly correlated.
In practice, however, this behaviour is constrained by
material properties, i.e., thermal conductivities and capacitances both within and across the boundaries of chips~\cite{hutter14}.
In other words, a linear correlation of (secret) activity patterns
and (leaked) thermal patterns is only applicable for homogeneous material
properties.
This has direct
implications for our work, as elaborated
in Section~\ref{sec:initial_findings}.
\item Due to
relatively large thermal capacitances and resistances present in modern ICs,
	   especially in 3D ICs,
 the internal heat flow is much slower than the underlying activity/power patterns (Figure~\ref{fig:power-thermal-scales}). That is, the TSC has a relatively low
bandwidth, which hinders the leakage of highly dynamic computation~\cite{hutter14}.
   We assume strong capabilities for practical attacks
(Section~\ref{sec:attacks}), rendering
	the TSC attractive nevertheless.
\item As with any side channel, the TSC experiences noisy readings; the heating effects of active modules are both spatially and temporally
superposed. For any
2D/3D IC, these noise patterns may range from negligible to dominant effects, depending on the floorplan, the resulting power distribution, the material properties, and the heat paths along with their
dissipation capabilities. Our
assumptions for the attacker's capabilities address and mitigate also this limitation (Section~\ref{sec:attacks}).
\end{compactitem}

\subsection{Scope of Our Work}

In this novel work, we focus on the design-time mitigation of thermal-related SCAs in 3D ICs.
In contrast, previous work leverages 3D ICs only as
implementation option for available (2D) SCA countermeasures, or neglects key aspects of 3D ICs (Section~\ref{sec:introduction}).

{\bf Our key idea} is the following: we exploit the specifics of material and structural properties in 3D ICs, in order to effectively decorrelate thermal patterns from the
power and activity patterns.

{\bf We conduct our study
	in the context of 3D floorplanning} for two reasons.
First, optimizing the thermal and power distributions of 3D ICs is essential (due to their high integration densities), and floorplanning is widely acknowledged as an appropriate
design stage to do so~\cite{Lim13,
		knechtel16_Challenges_ISPD,
	budhathoki16}.
Second,
chip designer typically have to reuse some ``black box'' intellectual-property (IP) modules, with limited access to only basic properties (e.g., area,
		inputs/outputs, power), while they still hope to mitigate potential SCAs.
Our methodology is applicable in such scenarios through leakage-aware
design
techniques during block-level floorplanning.

{\bf We make our work publicly available} within the open-source 3D floorplanner {\em Corblivar}~\cite{knechtel_CorblivarCode}; see Sections~\ref{sec:methodology}
and~\ref{sec:experiments} for details.

\section{Initial Findings}
\label{sec:initial_findings}

We next discuss 
	exploratory experiments for through-silicon via (TSV)-based 3D ICs
with two dies.\footnote{Thermal maps would be considerably different for other 3D integration flavors, e.g., for
	{\em monolithic 3D ICs}~\cite{samal16}. Our
leakage-mitigation techniques are generic; they may be tailored towards different flavors, which will be the scope for future work.}
For the thermal analysis we use {\em HotSpot 6.0}~\cite{zhang15}. We
	model the heatsink atop the 3D IC, the signal TSVs acting as ``heat-pipes'' between stacked dies, and also the secondary path
		conducting heat towards the package. Further details are given in Section~\ref{sec:experiments}.
We investigated all 30 combinations of 5 different power distributions
(globally uniform, locally uniform, medium gradients, small gradients, and large gradients) and 6 different TSV distributions (no TSVs; maximal TSV density, i.e.,
100\% of area covered by TSVs and their keep-out zones; irregular TSVs; irregular TSVs along with regular TSVs; irregular groups of densely packed TSVs, i.e., TSV islands; and TSV
islands along with regular TSVs). Note that some of these power and TSV distributions are impractical, yet relevant for exploratory experiments.
Figure~\ref{fig:initial-findings} illustrates a few selected power and thermal maps.

{\bf The key initial finding} for thermal-leakage trends in 3D ICs is that
the correlation of activity/power patterns and the thermal behaviour mainly depends
on both
($i$)~the power-density distributions and ($ii$)~the TSV distributions.
{\em We observe high correlations for:}

\vspace{1.25mm}
\begin{compactenum}

\item[($i$)]
    {\em non-uniform power distributions with large gradients, both within dies and across dies.}
For example,
Figure~\ref{fig:initial-findings}(e--h) exhibits
large power gradients, while
Figures~\ref{fig:initial-findings}(a--d) and \ref{fig:initial-findings}(i--l) show somewhat smooth
gradients.
We observe the lowest correlations for globally uniform power distributions
(Figure~\ref{fig:initial-findings}[a--d]), which are, however, artificial and impractical.
{\em Locally uniform}
power distributions are more realistic, and
we observe that such distributions exhibit low correlations as
well, especially along with irregular TSVs or distributed TSV islands.
For example, comparing
Figure~\ref{fig:initial-findings}(k, l) with
Figure~\ref{fig:initial-findings}(c, d), we observe largely decorrelated thermal patterns,
	although the
	underlying power distribution of the former
	appears notably more diverse than its counterpart.

\vspace{1mm}
\item[($ii$)]
{\em large numbers of regularly arranged TSVs}.
For example,
    see Figure~\ref{fig:initial-findings}(e--h) with regularly placed TSVs and TSV islands
	versus
Figures~\ref{fig:initial-findings}(a--d)
with irregularly placed TSVs
and \ref{fig:initial-findings}(i--l) with only TSV islands.
Note that the use of few regular TSVs and/or TSV islands may locally further reduce the correlation.
For example, see Figure~\ref{fig:initial-findings}(j), where a local thermal minima is observable in the lower right region, despite the relatively high power density underlying
(Figure~\ref{fig:initial-findings}[i]).

It may be more intuitive to formulate these findings vice versa---the less regular and/or fewer the TSVs, the lower is the correlation. As indicated for the
limitations of the TSC (Subsection~\ref{sec:background_TSC}), an unbiased
and high
correlation is only found in homogeneous structures. Inserting copper/tungsten-based TSVs into the
silicon dies invalidates
	that assumption, especially when TSVs are few and irregularly distributed.

\end{compactenum}

\section{Models for Thermal Leakage}
\label{sec:models}

\subsection{Correlation and Correlation Stability}

We propose the {\em Pearson correlation} of power and thermal maps as
a key metric for thermal-related leakage.
Note that the Pearson correlation is also the underlying measure for the {\em side-channel vulnerability factor (SVF)}~\cite{demme12}, an established metric for the
vulnerability of ICs in terms of information leakage via side channels.

Along with our assumptions for the attacker's capabilities (Section~\ref{sec:attacks}),
the Pearson correlation is comparably meaningful as the SVF.
That is, the lower the correlation, the lower the leakage of power/activity
patterns via the TSC, and the lower the vulnerability of the IC.
Given the power and thermal maps of a 3D IC, we
measure the
{\em correlation coefficient} $r_d$ on each die $d$
separately:
\begin{equation}
\label{eq:Pearson}
r_d = \frac{
\sum_{i=1}^{n} (p_i - \bar{p})(t_i - \bar{t})
}{
\sqrt{\sum_{i=1}^{n} \left(p_i - \bar{p}\right)^2}
\sqrt{\sum_{i=1}^{n} \left(t_i - \bar{t}\right)^2}
}
\end{equation}
where $\bar{p},\bar{t}$ are the average power and temperature values, respectively, and $p_i,t_i$ are the individual
values, both over all $n$
locations in die $d$.  The locations/values are to be organized in grids with same dimensions for both power and thermal maps/grids.

Note that Equation~\ref{eq:Pearson} models the correlation only for one {\em steady-state case}, i.e., for arbitrary but fixed power and temperature values. We propose another 
correlation measure to capture the {\em runtime stability of correlation} $r_{d,x,y}$ for locations/bins $x,y$ on die $d$:
\begin{equation}
\label{eq:corr_stability}
r_{d,x,y} = \frac{
\sum_{i=1}^{m} (p_{i,x,y} - \overline{p_{x,y}})(t_{i,x,y} - \overline{t_{x,y}})
}{
\sqrt{\sum_{i=1}^{m} \left(p_{i,x,y} - \overline{p_{x,y}}\right)^2}
\sqrt{\sum_{i=1}^{m} \left(t_{i,x,y} - \overline{t_{x,y}}\right)^2}
}
\end{equation}
with $m$ different sets of activities, where power and temperature readings may vary notably. The lower the
stability
$r_{d,x,y}$,
	the lower the thermal leakage for {\em various ranges} of power/activity patterns at particular locations $x,y$ within the die $d$ of the 3D IC.

\begin{figure*}[t]
\centering
\sidesubfloat[]{\includegraphics[width=.17\textwidth]{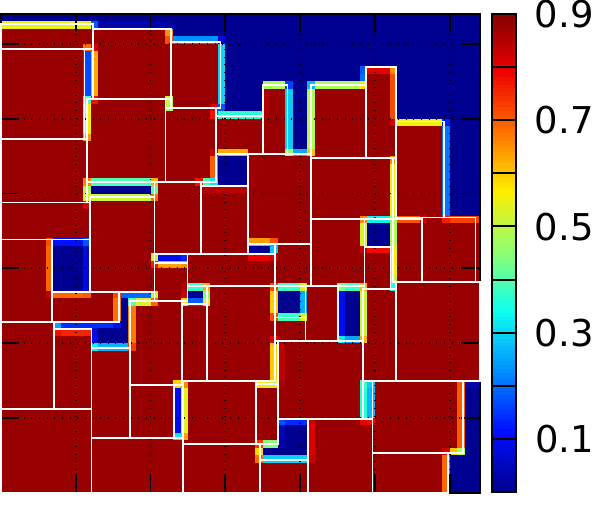}}\qquad
\sidesubfloat[]{\includegraphics[width=.17\textwidth]{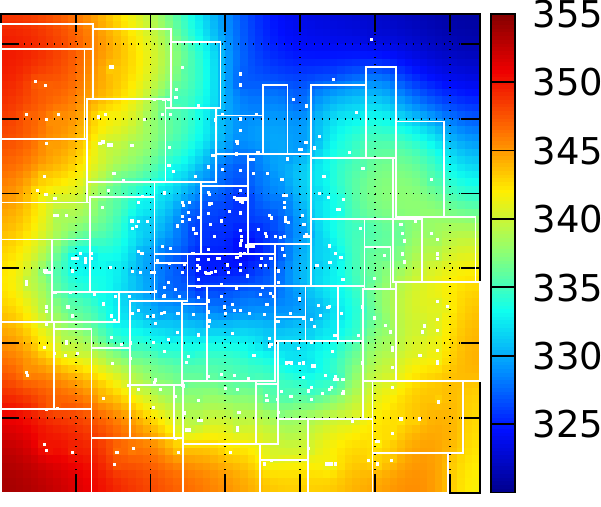}}\qquad\qquad
\sidesubfloat[]{\includegraphics[width=.17\textwidth]{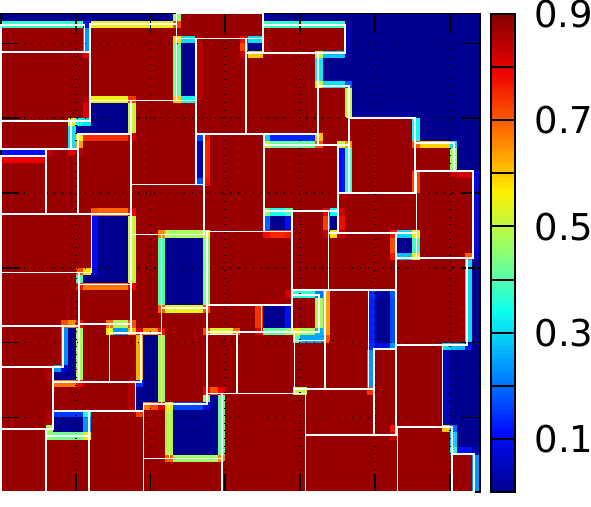}}\qquad
\sidesubfloat[]{\includegraphics[width=.17\textwidth]{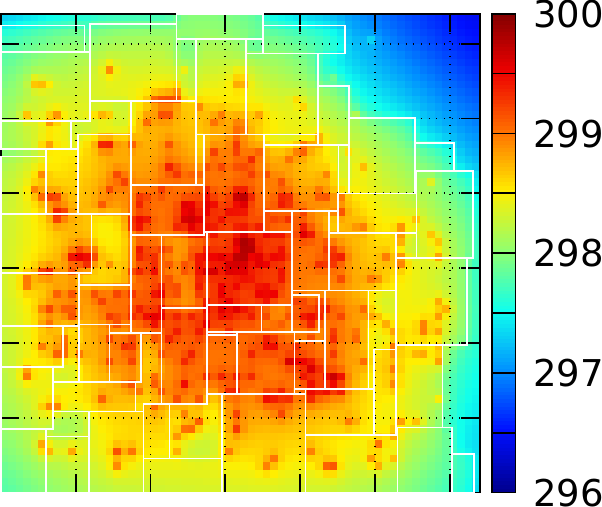}}\\
	\vspace{0.3em}
\sidesubfloat[]{\includegraphics[width=.17\textwidth]{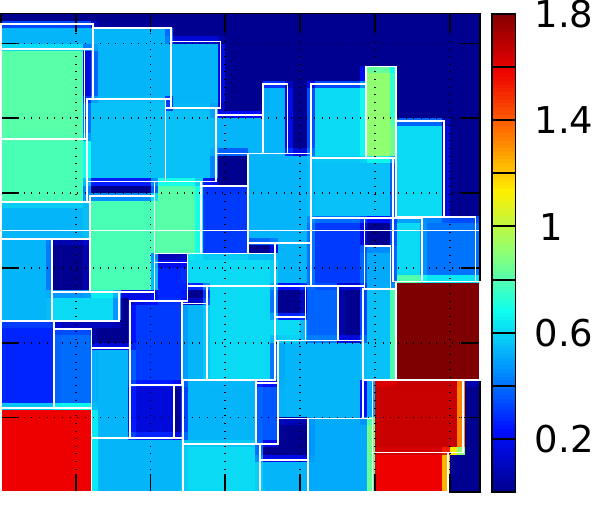}}\qquad
\sidesubfloat[]{\includegraphics[width=.17\textwidth]{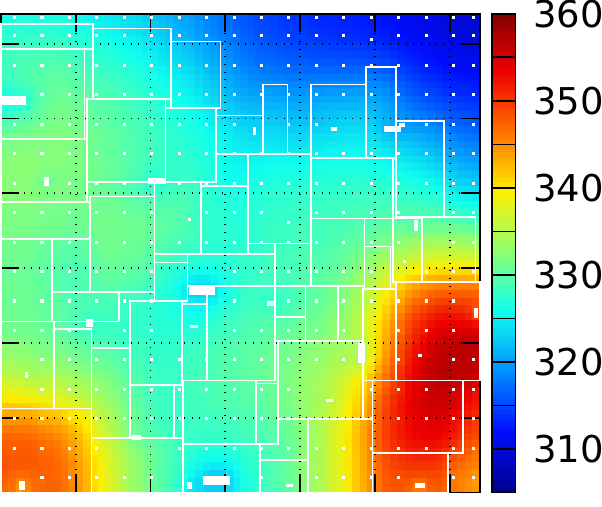}}\qquad\qquad
\sidesubfloat[]{\includegraphics[width=.17\textwidth]{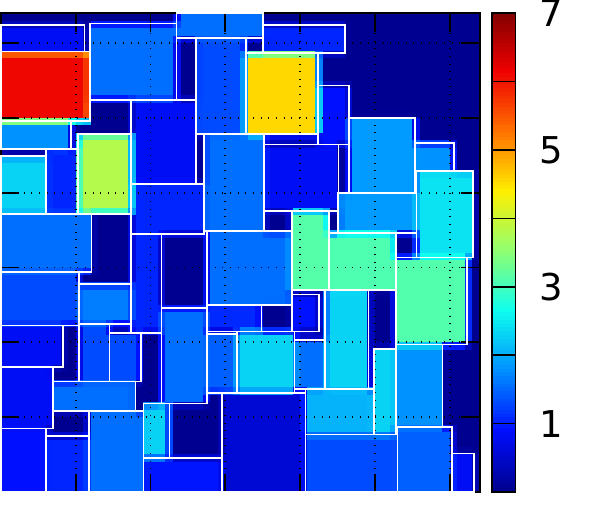}}\qquad
\sidesubfloat[]{\includegraphics[width=.17\textwidth]{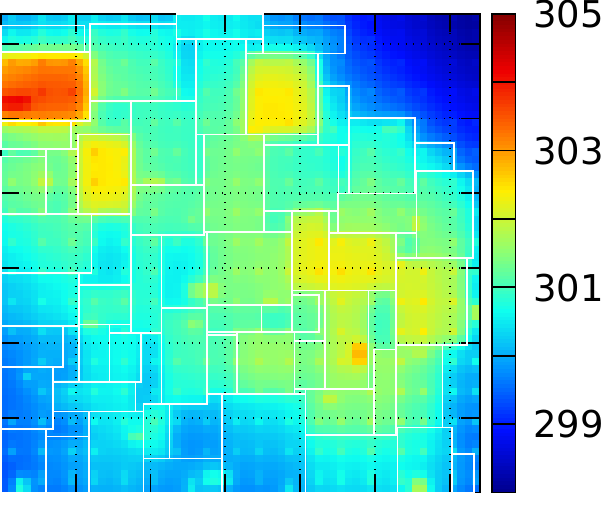}}\\
	\vspace{0.3em}
\sidesubfloat[]{\includegraphics[width=.17\textwidth]{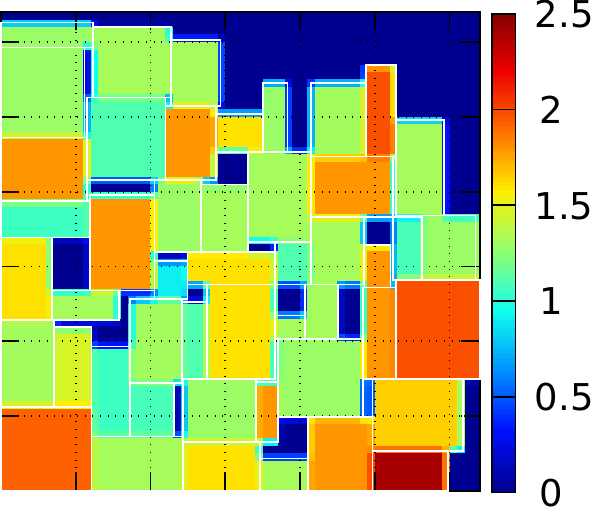}}\qquad
\sidesubfloat[]{\includegraphics[width=.17\textwidth]{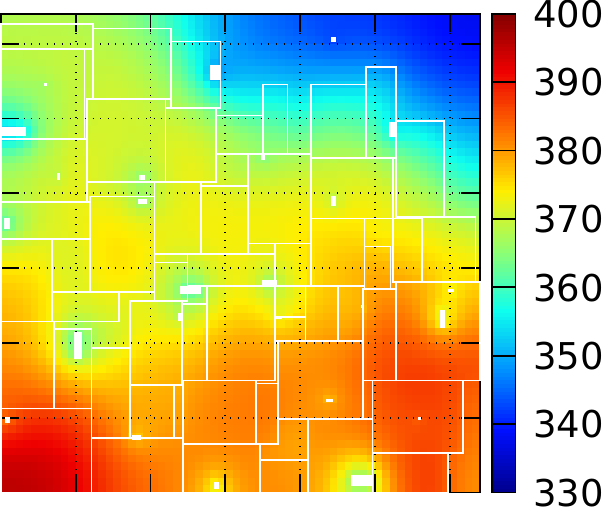}}\qquad\qquad
\sidesubfloat[]{\includegraphics[width=.17\textwidth]{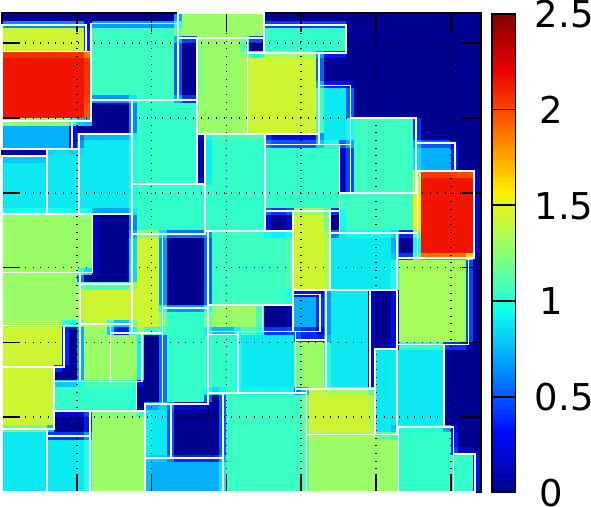}}\qquad
\sidesubfloat[]{\includegraphics[width=.17\textwidth]{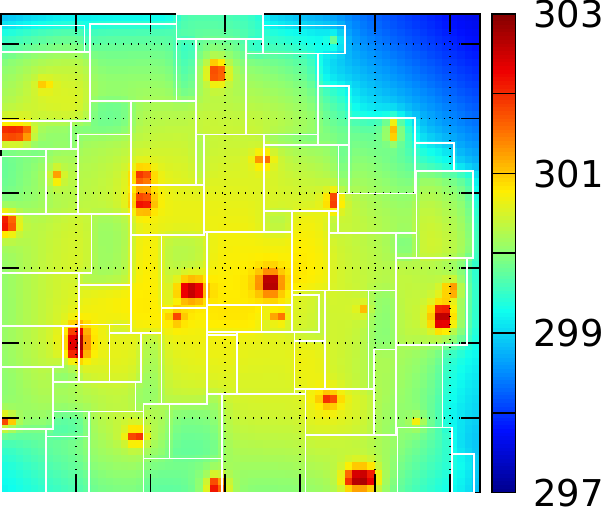}}
	\smallerspacecaption
	\caption{A 3D IC floorplanned on two dies, with three different power scenarios illustrated in rows.  From left to right, each row illustrates the power-density map of
		the lower die (a, e, i), the thermal map for that die (b, f, j), and the power (c, g, k) and thermal map (d, h, l) for the upper die.
			Each scenario exhibits distinct patterns for power and TSV distributions (the latter are illustrated as white dots in [b, f, j]), resulting in different
			trends for the power-temperature correlations
			as follows.
		Top row:
			artificially unified power for all modules with irregularly
			placed TSVs,
			resulting in the lowest correlation;
		middle row:
			large power gradients with
			both regular TSVs and TSV islands,
			resulting in the highest correlation;
		bottom row: groups of locally similar power regimes with 
		       TSV islands,
		       resulting in relatively low correlation.
			The units of the power and thermal maps are $10^{-2} \mu W / \mu m^2$ and $K$, respectively.
    \label{fig:initial-findings}
	}
\end{figure*}

\subsection{Spatial Entropy of Power Maps}

As proposed by Claramunt~\cite{claramunt05}, the spatial entropy
assesses
the dispersion of the classical entropy over some regions.
It is based on two principles:
($i$) the closer the {\em different} entities, the {\em higher} the spatial entropy; and ($ii$) the closer the {\em similar} entities, the {\em lower} the spatial entropy.
We note that this is in good correspondence to the
phenomenon of heat distribution---the closer the {\em differently powered} heat sources, the {\em higher}
    the spatial entropy of the superposed thermal responses, i.e.,
the thermal gradients; and
	the closer the {\em similarly powered} heat sources, the {\em lower} the thermal gradients.

Hence, our notion for modeling the spatial entropy of power maps is to anticipate
the degree of the thermal gradients;
recall that large thermal gradients hint on large thermal leakage (Section~\ref{sec:initial_findings}).
Our
formulation,
derived from~\cite{claramunt05},
measures the spatial entropy of power maps in 3D ICs for each die $d$
	as follows:
\begin{equation}
S_d = - \sum_{i=1}^{n} \frac{d_i^{inter}}{d_i^{intra}} \left( \frac{|c_i|}{|C|} \log_2 \frac{|c_i|}{|C|} \right)
\end{equation}
where $c_i \in C$ are {\em classes of similar-value power ranges},
and $d_i^{inter}$ and $d_i^{intra}$ are the
average spatial {\em inter- and intra-class distances}.
    We calculate those distances
according to~\cite{claramunt05}, in which
we use Manhattan distances for
  bins (i.e., class members)
   in the equidistant power-map
   grids we construct for each die.
For fast and effective classification of those grids,
	we employ
{\em nested-means partitioning}: the power values are first sorted, then recursively bi-partitioned with the current
mean defining the cut,
     and the partitioning proceeds
     until
     the standard deviation within any class approaches zero.

As the spatial entropy does not account for thermal analysis and subsequent correlation calculations,
it lacks the capability to verify the actual leakage. It is, however, suitable for fast estimation of the potential leakage during floorplanning loops/iterations.
During our experiments (Sections~\ref{sec:initial_findings} and~\ref{sec:experiments}), we observe the following trend for the bottom die ($d=1$),
    even for different TSV patterns: the lower the spatial entropy, the
lower the power-temperature correlation.

\section{Thermal Side-Channel Attacks}
\label{sec:attacks}

{\bf Our assumptions:}
An attacker has direct and physical access to the targeted 3D IC, but can conduct only non-invasive attacks.\footnote{In contrast to 2D ICs, we may expect invasive probing and
	reverse engineering of 3D ICs to be notably more difficult and costly.}
That is, he/she may apply (arbitrary) input patterns and observe both the actual outputs and the thermal
behaviour of the 3D IC.\linebreak
We also assume that the attacker has a system-level
understanding of the 3D IC, e.g., as obtained from datasheets. This is important
for purposefully crafting input patterns
to trigger certain activities.

We make further strong
	assumptions enabling an attacker to circumvent the critical limitations of the TSC
(Subsection~\ref{sec:background_TSC}).
First, the attacker can
stabilize the 3D IC's
activity
	with the help of specifically crafted, repetitive input patterns.
Second, he/she may
await the thermal steady-state response after applying any input.
Both assumptions
ensure that the TSC readout correlates
only with the
input patterns.
Third,
the attacker has unlimited access to all thermal sensors, spread across
the 3D IC, and can thus obtain high-accuracy and continuous thermal readings of
any (part of a) module at will.

We believe that these assumptions are relevant and practical when attacking security-critical (3D) ICs. For example, a security module may check whether a
provided password is correct, and only then trigger data decryption.
The thermal patterns for complex decryption operations will be relatively easy to distinguish from simple matching operations for password checks.
   An attacker may then brute-force a password even when the security module itself
is ``silent'', i.e., when it would not directly react to wrong passwords.

{\bf Attacks based on the TSC in 3D ICs} are formulated next, with consideration of the attacker's capabilities outlined above.
Note that similar attacks have been discussed and successfully conducted on classical (2D) microcontrollers by Hutter and Schmidt~\cite{hutter14}.

\vspace{1.25mm}
\begin{compactenum}

\item {\em Thermal characterization of the 3D IC:}
This is typically an exploratory attack, and others may follow as outlined below.
Step by step, the attacker will apply
a broad and
varied range of input patterns in order to trigger as many activity patterns as possible. By monitoring the TSC,
	 he/she can then build a model for the thermal behaviour
of the 3D IC.

\item {\em Localization and monitoring of modules:}
The attacker targets on particular
modules by applying
crafted input patterns; the objective is
to trigger
these modules and 
observe thermal variations exclusively or at least predominantly within these modules.
Given the attacker's lack of implementation details and hence the lack of a precise input-to-activity mapping, this attack is typically applied iteratively, with varying
inputs.
In case the characterization attack outlined above has been successfully applied, those efforts may be 
much lower.
Once the thermal response is confined to particular regions, i.e., modules of interest are localized with some confidence, further attacks
may be applied. Most notably, an attacker may now observe the sensitive activity/computation of particular modules by monitoring them during runtime.

\end{compactenum}

\section{Methodology}
\label{sec:methodology}

\begin{figure}[t]
\centering
	\includegraphics[width=.8\columnwidth]{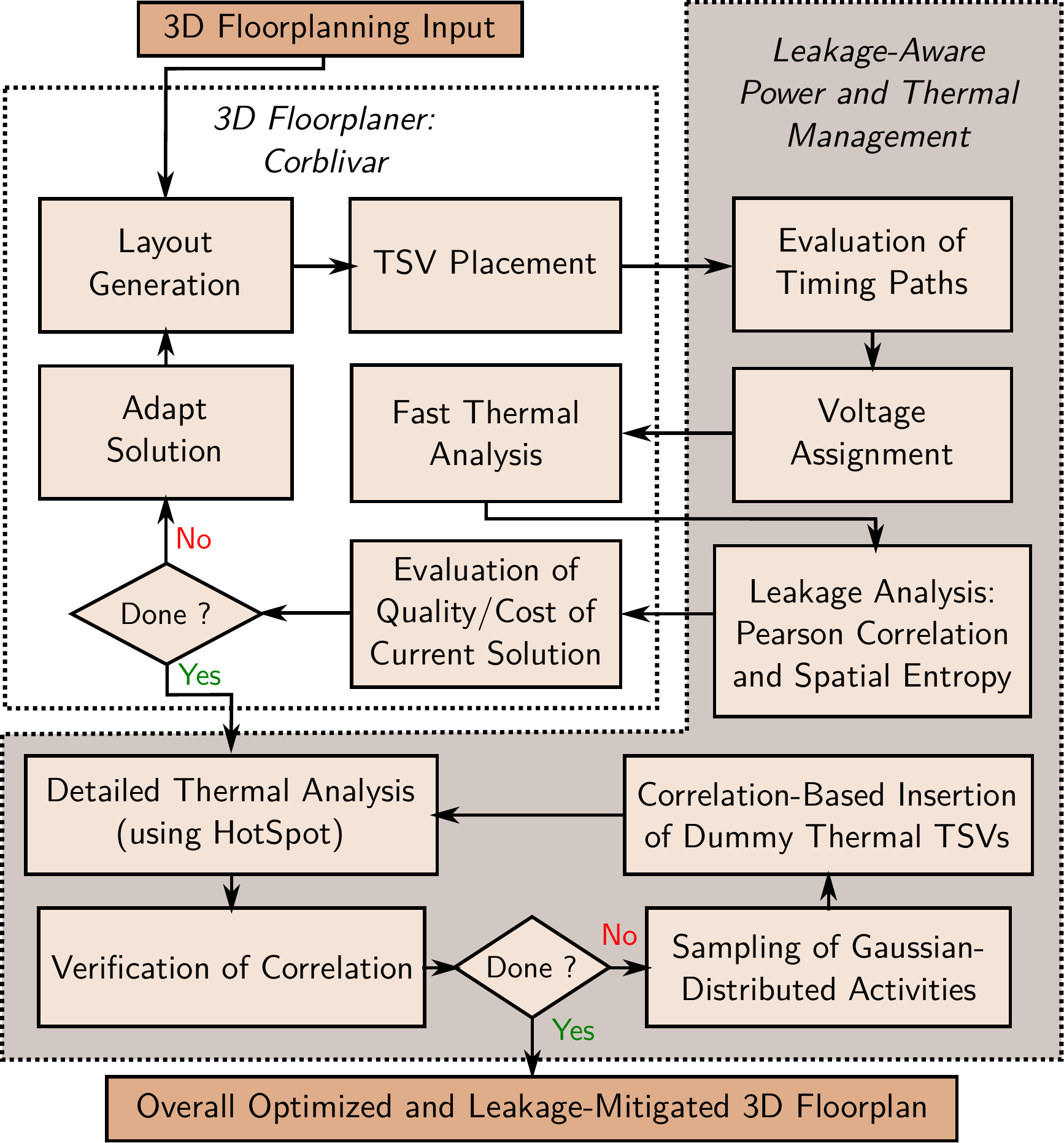}
	\smallerspacecaption
	\caption{The flow of our methodology, implemented as extension for the open-source 3D floorplanner {\em Corblivar}~\cite{knechtel_CorblivarCode, knechtel15_Corblivar_TCAD}.
    \label{fig:flow}
	}
\end{figure}

The objective of our work is to account for thermal leakage in 3D ICs {\em directly and continuously within the floorplanning stage} that also optimizes
for
classical criteria.
That is, during our iterative floorplanning flow (Figure~\ref{fig:flow}), we memorize the 3D floorplans with lowest correlations coefficients and spatial entropies (Section~\ref{sec:models})
as best in terms of thermal leakage, but we select the final
solution while also accounting for 
other criteria such as wirelength and critical delays; see Section~\ref{sec:experiments} for details of our setups.

We
	implement and publicly release all our techniques
within {\em Corblivar}~\cite{knechtel15_Corblivar_TCAD, knechtel_CorblivarCode}. We opt for this 3D floorplanning framework mainly because it is
multi-objective, modular, and competitive.
An for our work essential feature of {\em Corblivar} is its fast thermal analysis,
which enables us to continuously estimate the correlation.
However, we found this fast analysis to be inferior to the detailed analysis of {\em HotSpot}~\cite{zhang15},
especially for diverse arrangements of TSVs.
Thus, we also verify the final correlation after floorplanning.

\subsection{Management of Power Distributions}

Recall that non-uniform power distributions with large gradients induce notable thermal leakage (Section~\ref{sec:initial_findings}).
A key measure in our work is thus the
	management of global and local power distributions.
	To this end, we develop 
efficient algorithms for floorplanning-centric {\em voltage assignment}, a technique
well-known for power management in 2D ICs~\cite{ma11_MSV}, but so far mostly overlooked for 3D ICs.

	For any floorplan layout, we initially
	estimate the timing paths.
We do so because
the prospects for voltage assignment depend primarily on timing slacks---the more slack a module has, the lower the voltage we {\em may} apply, and
the more controllable its power and thermal footprints become.
We estimate the net delays via the well-known {\em Elmore delays} (here with consideration of wires and TSVs),
and the delays of
modules are estimated as proposed in~\cite{lin10}.

{\em Voltage volumes}---the generalized 3D version of voltage domains spanning across multiple dies---are constructed next.
We do so by considering each module individually as the root for a {\em multi-branch tree} representation of voltage volumes.
Each tree/volume is recursively build up via a breadth-first search across the respectively adjacent modules.
During this merging procedure, we
update the resulting set of {\em feasible voltages}, i.e., the voltages we may commonly apply for all modules within the
volume without violating any timing/delay constraints.
As result, a finalized tree represents a set of practical voltage volumes: each node comprises a
volume,
	and all the modules and voltages are encoded in the node and its ancestors.

Finally, we select voltage volumes such that we optimize
($i$) for locally uniform power densities within volumes, and ($ii$) for small power gradients across volumes, as motivated by our initial findings (Section~\ref{sec:initial_findings}).
To this end, we target for lowest
standard deviations of power values both within voltage volumes and across volumes.

We found that our techniques induce a low runtime cost, around 30\%, when compared to 3D floorplanning without voltage
assignment. This is
noteworthy
	     because previous work~\cite{lin10,lee14} employ computationally-expensive {\em MILP formulations} which
render their integration into 3D floorplanning flows
	impractical in the first place.

\subsection{Sampling of Activities and Post-processing}

To impersonate an attacker triggering various activity
patterns by alternating the inputs at runtime,
we model the power profiles of all modules as Gaussian distributions.
Without loss of generality, we set up each distribution
with the module's nominal power value as mean and a standard deviation of 10\%.
We stepwise evaluate all the steady-state temperatures using {\em HotSpot}~\cite{zhang15} and sample the correlation stability (Equation~\ref{eq:corr_stability}) in 100 runs over the whole 3D IC.

Continuing the runtime sampling process, we iteratively insert {\em dummy thermal TSVs} where
the most stable correlations occur, as long as the resulting
average correlation is reduced. This stop criterion represents the final ``sweet spot'' where further TSV insertion would increase the overall correlation again
(Subsection~\ref{sec:destabilizing}).

\section{Experimental Evaluation}
\label{sec:experiments}

We select arbitrary {\em GSRC} and {\em IBM-HB+} benchmarks, and
their properties are reviewed in Table~\ref{tab:benchmarks}. Note that we scale up the modules' footprints in order to obtain sufficiently large dies; 3D ICs are only
superior to 2D ICs for large-scale integration~\cite{knechtel16_Challenges_ISPD, knechtel15_Corblivar_TCAD, Lim13}.
	The resulting die outlines are fixed, making the floorplanning problem practical yet challenging~\cite{knechtel15_Corblivar_TCAD}.
For stacking of the two dies,
we assume
the {\em face-to-back} fashion~\cite{Lim13,ITRS15} (see also Figure~\ref{fig:power-thermal-scales}).
Further technical details such as material properties and TSV dimensions
	  are given in the respective default configurations of~\cite{zhang15,knechtel_CorblivarCode}.

Each benchmark is floorplanned 50 times in two different setups: ($i$)~power-aware floorplannning and ($ii$)~TSC-aware
floorplanning.
		It is important to note that
setup ($i$) provides a competitive and challenging baseline for the evaluation of ($ii$).
Unlike we do in ($i$),
	most previous work did not account for continuous optimization of the voltage assignment within the
optimization loops, and may have thus missed on the full potential of power-aware 3D floorplanning.

For ($i$),
    we optimize the packing density, wirelength, critical delay, peak temperature, and voltage assignment, all at the same time; all criteria are weighted equally.
	For voltage assignment we
	seek to
	minimize both the overall power and the number of required voltage volumes.
For ($ii$), we consider the same criteria as for ($i$), rendering it a practical and competitive setup.
Additionally, we
seek to minimize both the average correlation
coefficients and the average spatial entropies.
Here, for voltage assignment, we now seek to minimize ($a$) the number of required voltage volumes and ($b$) the standard deviations of power gradients among and across different volumes.
For implementing the voltage volumes in both setups ($i$) and ($ii$), we consider three options:
0.8V, with power scaling of 0.817$\times$ and delay scaling of 1.56$\times$; 1.0V, without impact on power/delay;
	and 1.2V, with 1.496$\times$ power scaling and 0.83$\times$ delay scaling. These values are simulated for the 90nm node~\cite{lin10}.

\begin{table}[b]
\centering
\scriptsize
\setlength{\tabcolsep}{1mm}
\renewcommand{\arraystretch}{1.1}
\begin{tabular}{|c||c|c|c|c|c|c|}
\hline

Name &
\# Modules &
Modules'&
\# Nets &
\# Terminal &
Outline &
Power (for\\

&
(Hard/Soft) &
Scale Factor &
&
Pins &
[$mm^2$] &
1.0V) [W] \\

\Hline

{\em n100} &
(0/100) &
10 &
885 &
334 &
16 &
7.83
\\

\rowcolor{gray}
{\em n200} &
(0/200) &
10 &
1,585 &
564 &
16 &
7.84
\\

{\em n300} &
(0/300) &
10 &
1,893 &
569 &
23.04 &
13.05
\\

\hline

{\em ibm01} &
(246/665) &
2 &
5,829 &
246 &
25 &
4.02
\\

\rowcolor{gray}
{\em ibm03} &
(290/999) &
2 &
10,279 &
283 &
64 &
19.78
\\

{\em ibm07} &
(291/829) &
2 &
15,047 &
287 &
64 &
9.92
\\

\Hline

\end{tabular}

\smallerspacecaption
\caption{
	Properties of {\em GSRC} and {\em IBM-HB+} Benchmarks
\label{tab:benchmarks}
}
\end{table}

\subsection{On Destabilizing the Leakage Correlation}
\label{sec:destabilizing}

A low runtime stability of correlations is crucial to thwart TSC attacks (Section~\ref{sec:attacks}):
the lower the correlation for various inputs, the less likely an attacker succeeds when modeling the thermal leakage.
We observe that our techniques
greatly reduce both
the runtime stability of correlations and the nominal, steady-state correlations.

\begin{figure}[t]
\centering
\sidesubfloat[]{\includegraphics[width=.435\textwidth]{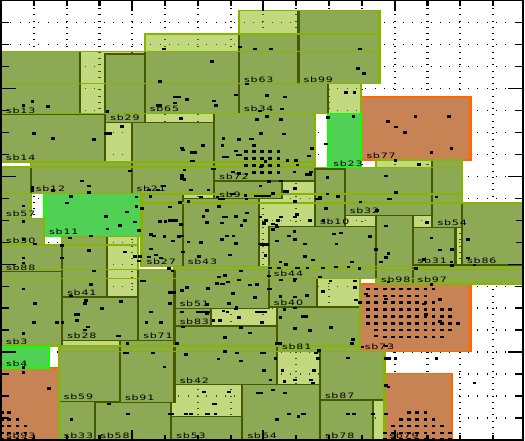}}\hfill
\sidesubfloat[]{\includegraphics[width=.435\textwidth]{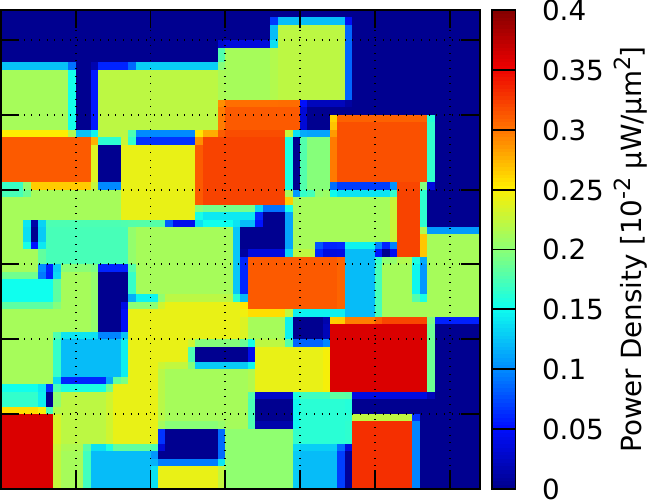}}\\
	\vspace{0.3em}
\sidesubfloat[]{\includegraphics[width=.435\textwidth]{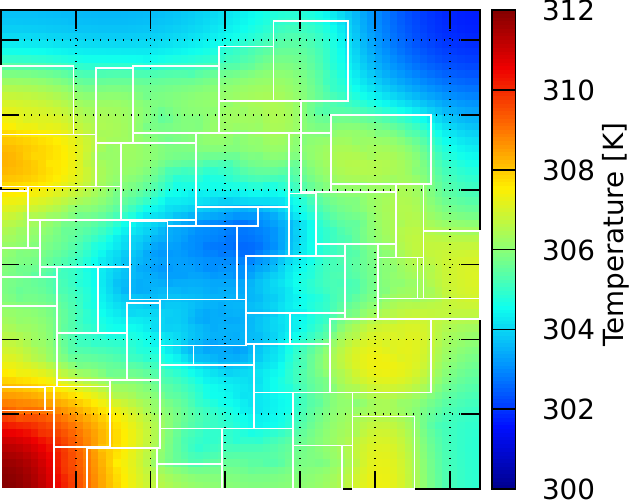}}\hfill
\sidesubfloat[]{\includegraphics[width=.435\textwidth]{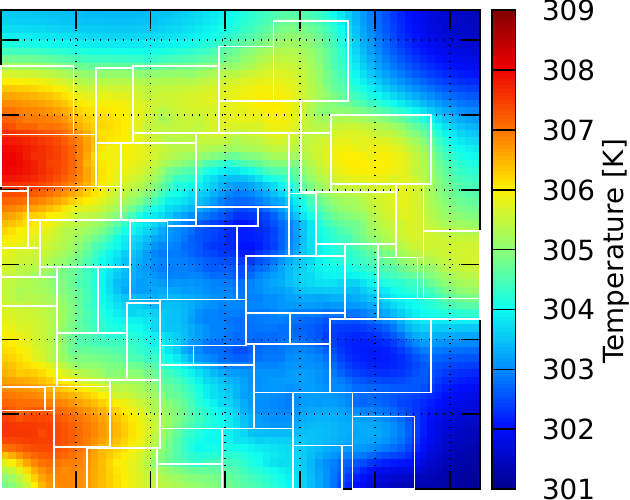}}\\
		\smallerspacecaption
	\caption{
	(a) An exemplary floorplan for the bottom die of benchmark {\em n100}, (b) its power distribution, the thermal maps (c) before and (d) after TSV insertion by post-processing.
    \label{fig:stability}
	}
\end{figure}

Consider
Figure~\ref{fig:stability} as an example. In order to meet timing constraints, some modules (colored red in Figure~\ref{fig:stability}[a])
had to have high voltages assigned. The accordingly high power consumption of those modules disrupts the otherwise more uniform power
distribution (Figure~\ref{fig:stability}[b]).
As a result, for varying activity patterns triggered by an attacker, the highest stability of correlations occurs around those modules. Our
post-processing stage consequently inserts dummy TSVs there (see black dots in Figure~\ref{fig:stability}[a]). For the nominal case, i.e., for average activity patterns, the 
resulting shift in thermal behaviour is reflected in
Figure~\ref{fig:stability}(c, d).
In this example, the correlation coefficient drops from 0.461 to 0.324.
This means that an attacker seeking to retrieve sensitive activities based on thermal
	readings is on average $\approx$~30\% less likely to succeed.

We find that
the insertion of dummy thermal TSVs may {\em stabilize} the correlation again, by inducing adverse side-effects on previously more decorrelated regions.
For example, consider the upper left corner in Figure~\ref{fig:stability}(c, d):
	the local correlation there increased to some degree, due to the insertion of TSVs in the other, previously highly correlated regions like the lower right corner.
To account for such trade-off effects, recall that our post-processing stage inserts dummy TSVs only as long as the overall correlation decreases.
Alternatively, we may adapt that stage to focus on reducing the correlation stability primarily for the critical module(s) to be protected from TSC attacks, and to accept
more stable correlations elsewhere.

\subsection{Leakage Trends and Mitigation Rates}

\begin{figure}[t]
\centering
	\includegraphics[width=.99\columnwidth]{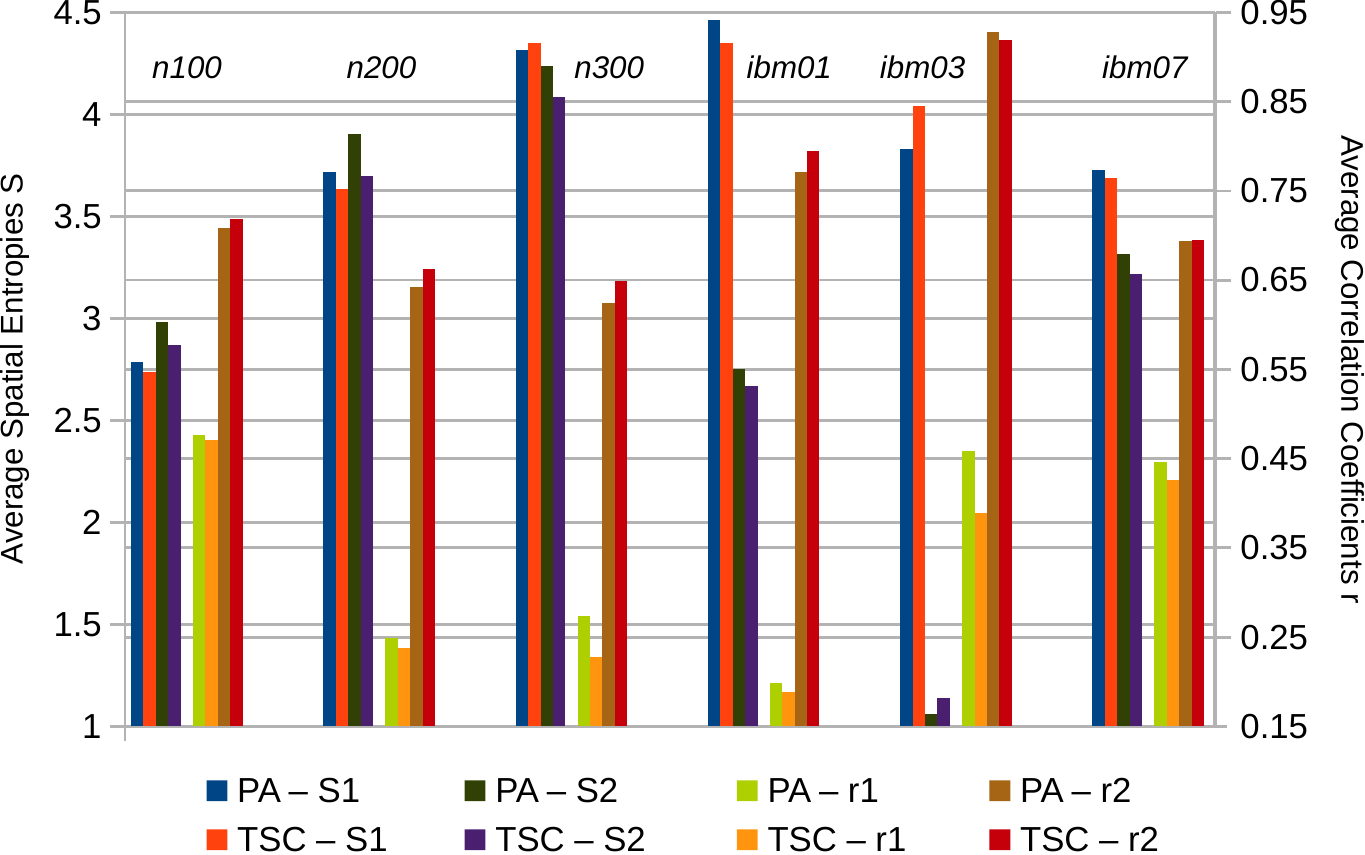}
	\smallerspacecaption
	\caption{Spatial entropies (S1, S2) and correlation coefficients (r1, r2) for power-aware floorplanning (PA) and thermal side-channel-aware floorplanning (TSC). Illustrated
		are the average ranges over 50 floorplanning runs; see also Table~\ref{tab:results-floorplanning}.
    \label{fig:results-steady-state}
	}
\end{figure}

To understand general trends on the thermal leakage in 3D ICs and the mitigation rates offered by our techniques, we next
discuss the average ranges of spatial entropies and correlation coefficients over 50 runs (Figure~\ref{fig:results-steady-state} and Table~\ref{tab:results-floorplanning}).
However, it is important to note that
these numbers do
not reflect well on best
      cases, which a designer will select carefully depending on the needs for security
      and the margin for design cost; see Subsection~\ref{sec:design-cost} for the latter.

When comparing the power-aware and the TSC-aware results, the key observation is that the latter exhibit
lower correlations for the bottom die ($r_1$),
by 16.79\% for the largest {\em GSRC} benchmark {\em n300},
      by 15.25\% for the largest {\em IBM-HB+} benchmark {\em ibm03},
      and
      by 7.71\% over all benchmarks on average.
While the average reduction over all benchmarks may appear somewhat low, its impact is important. It translates directly to 7.71\% higher noise on average for an attacker, and to an
	accordingly higher degree of freedom for a designer to employ sensitive modules.
	Besides, a noteworthy trend here is the scalability: the larger the circuit, the
	lower the correlation achieved by our TSC-aware floorplanning.

Both the power-aware and the TSC-aware setup can achieve low correlations for the bottom die, but at the cost of notably higher correlations for the top
die.
This limitation is due to the following: since the heatsink is attached above the top die, {\em Corblivar}~\cite{knechtel15_Corblivar_TCAD, knechtel_CorblivarCode} by
default employs a
thermal design rule to place high-power modules preferably into the top die. In turn, this results in large power gradients across the two dies as well as within
the top die.
Large gradients in the top die naturally increase its correlation.
Furthermore, the large disparities between the two dies obfuscate any ``heat injection'' from the bottom die into the top die which may otherwise decrease the top die's
correlation, e.g., see Figure~\ref{fig:initial-findings}(e--h) versus Figure~\ref{fig:initial-findings}(a--d).
While we may relax {\em Corblivar}'s design rule, we found that this prohibitively increases the peak temperatures.

\subsection{Design Impact and Effectiveness}
\label{sec:design-cost}

Next, we discuss the average impact on other design criteria (Table~\ref{tab:results-floorplanning}).
Again, these results are relevant to understand general trends, while they do not represent particular corner cases.

Recall that a key measure in our methodology is the management of global and local power distributions, realized by tailored voltage assignment.
Our technique increases power only by 5.38\% on average when compared to power-aware floorplanning, whose key criterion is minimal power.
Despite this (limited) increase of power, we observe notable reductions of the steady-state peak temperatures, namely by 13.22\% on average (i.e., with respect to the ambient temperature of 293K).
We consider this as a beneficial side-effect of our multi-objective and iterative 3D floorplanning flow.

Besides,
we observe an average impact on other criteria as follows:
wires are 1.08\% longer,
     critical delays increase by 10.33\%,
	and, most notably,
     87.17\% more voltage volumes are implemented.

Finally, we note that ($i$) power- and TSC-aware floorplanning exhibit the same average number of signal TSVs, and that ($ii$) TSC-aware floorplanning employs only few
additional thermal TSVs (1.37\% on average).
Both findings indicate that our TSC-aware floorplanning realizes the reported low correlation
not only by carefully inserting dummy TSVs, but more so by thoroughly exploring the 3D design space. This is achieved for the first time thanks to our effective and efficient
floorplanning
techniques.

\begin{table*}[t]
\vspace{-1.22em}
\centering
\scriptsize
\setlength{\tabcolsep}{1.1mm}
\renewcommand{\arraystretch}{1.02}
\begin{tabular}{|c||c|c|c|c|c|c|c||c|c|c|c|c|c|c|}

\hline

& \multicolumn{7}{c||}{Power-Aware Floorplanning}
& \multicolumn{7}{c|}{Thermal Side-Channel-Aware Floorplanning} \\
\cline{2-15}

Metric
& \multicolumn{3}{c|}{GSRC Benchmarks}
& \multicolumn{3}{c|}{IBM-HB+ Benchmarks}
&
& \multicolumn{3}{c|}{GSRC Benchmarks}
& \multicolumn{3}{c|}{IBM-HB+ Benchmarks}
&	\\
\cline{2-7}
\cline{9-14}

& {\em n100} & {\em n200} & {\em n300}
& {\em ibm01} & {\em ibm03} & {\em ibm07} & {Avg}
& {\em n100} & {\em n200} & {\em n300}
& {\em ibm01} & {\em ibm03} & {\em ibm07} & {Avg}	\\
\Hline

			Spatial Entropy $S_1$
&2.787	&3.718	&4.315
&4.462	&3.830	&3.727
&3.806

&2.735	&3.635	&4.350
&4.348	&4.040	&3.689
&3.799	\\

\rowcolor{gray}		Correlation Coefficient $r_1$
&0.476	&0.249	&0.274
&0.199	&0.459	&0.446
&0.351

&0.471	&0.238	&0.228
&0.189	&0.389	&0.426
&0.324	\\

			Spatial Entropy $S_2$
&2.982	&3.905	&4.237
&2.752	&1.061	&3.314
&3.041

&2.868	&3.698	&4.086
&2.668	&1.140	&3.219
&2.946	\\

\rowcolor{gray}		Correlation Coefficient $r_2$
&0.708	&0.642	&0.625
&0.771	&0.928	&0.694
&0.728

&0.719	&0.662	&0.649
&0.795	&0.919	&0.695
&0.739	\\

\hline

			Overall Power [W]
&7.890	&7.801	&12.522
&4.291	&27.038	&10.737
&11.713

&10.331	&8.172	&12.967
&4.520	&26.829	&11.247
&12.344	\\

\rowcolor{gray}		Critical Delay $[ns]$
&0.833	&0.784	&1.164
&1.49	&3.54	&2.82
&1.771

&0.878	&0.817	&1.162
&1.607	&3.831	&3.434
&1.954	\\

			Wirelength
			$[m]$
&30.671	&29.086	&37.100
&21.491	&57.999	&108.020
&47.394

&32.956	&28.733	&33.928
&21.913	&57.449	&112.468
&47.907	\\

\rowcolor{gray}		Peak Temp $[K]$~\cite{zhang15}
&309.811	&308.475	&311.095
&303.496	&334.759	&308.715
&312.725

&315.563	&309.246	&312.168
&303.459	&311.066	&309.201
&310.117	\\

			Signal TSVs
&451	&897	&1,100
&3,489	&3,809	&10,740
&3,414

&451	&897	&1,099
&3,490	&3,806	&10,741
&3,414	\\

\rowcolor{gray}		Dummy Thermal TSVs
&--	&--	&--
&--	&--	&--
&--

&64	&29	&25
&3	&16	&144
&47	\\

			Voltage Volumes
&6.900	&12.365	&14.365
&3.675	&5.333	&3.027
&7.610

&9.705	&17.048	&19.658
&14.256	&12.909	&11.888
&14.244	\\

\rowcolor{gray}		Runtime [s]
&83	&218	&262
&489	&197	&507
&226

&287	&531	&524
&1075	&374	&667
&576	\\

\Hline

\end{tabular}

\caption{
	Average Spatial Entropies, Correlation Coefficients (Top) and Average Design Cost (Bottom) Over 50 Runs of Power-Aware Floorplanning (Left) versus Thermal
		Side-Channel-Aware Floorplanning (Right)
\label{tab:results-floorplanning}
}
\smallerspacecaption
\end{table*}

\section{Conclusion}

In this work, we address the thermal leakage of secret computation/activity patterns within up-and-coming 3D ICs.
In general, thermal side-channel attacks are
considered practical and effective; this is in contrast to other laborious attacks such as invasive probing, which are for 3D ICs
presumably even more difficult.

We model thermal leakage as a built-in criteria into floorplanning and minimize it using novel and efficient techniques, thereby incorporating security early on in the 3D IC design flow.
Our techniques are based on the key findings that thermal leakage in
3D ICs depends on and, thus, can be controlled by the power distributions along with the arrangement of through-silicon vias.

For future work, we shall address the thermal leakage in larger 3D-IC stacks and for other flavours such as monolithic integration.

\refsize
\newcommand{\BIBdecl}{\setlength{\itemsep}{0.18em}\vfill}
\bibliographystyle{IEEEtran}
\bibliography{main}

\begin{thebibliography}{10}
\providecommand{\url}[1]{#1}
\csname url@samestyle\endcsname
\providecommand{\newblock}{\relax}
\providecommand{\bibinfo}[2]{#2}
\providecommand{\BIBentrySTDinterwordspacing}{\spaceskip=0pt\relax}
\providecommand{\BIBentryALTinterwordstretchfactor}{4}
\providecommand{\BIBentryALTinterwordspacing}{\spaceskip=\fontdimen2\font plus
\BIBentryALTinterwordstretchfactor\fontdimen3\font minus
  \fontdimen4\font\relax}
\providecommand{\BIBforeignlanguage}[2]{{%
\expandafter\ifx\csname l@#1\endcsname\relax
\typeout{** WARNING: IEEEtran.bst: No hyphenation pattern has been}%
\typeout{** loaded for the language `#1'. Using the pattern for}%
\typeout{** the default language instead.}%
\else
\language=\csname l@#1\endcsname
\fi
#2}}
\providecommand{\BIBdecl}{\relax}
\BIBdecl

\bibitem{rostami14}
M.~Rostami \emph{et~al.}, ``A primer on hardware security: Models, methods, and
  metrics,'' \emph{Proc. IEEE}, vol. 102, no.~8, pp. 1283--1295, 2014.

\bibitem{zhou05}
Y.~Zhou and D.~Feng, ``Side-channel attacks: Ten years after its publication
  and the impacts on cryptographic module security testing,'' in \emph{IACR
  Crypt. ePrint Arch.}, no. 388, 2005.

\bibitem{skorobogatov12}
S.~Skorobogatov and C.~Woods, ``In the blink of an eye: There goes your {AES}
  key,'' in \emph{IACR Crypt. ePrint Arch.}, no. 296, 2012.

\bibitem{hutter14}
M.~Hutter and J.-M. Schmidt, ``The temperature side channel and heating fault
  attacks,'' in \emph{Smart Card Research and Advanced Applications}, ser.
  Lect. Notes Comp. Sci.\hskip 1em plus 0.5em minus 0.4em\relax Springer, 2014,
  vol. 8419, pp. 219--235.

\bibitem{masti15}
R.~J. Masti \emph{et~al.}, ``Thermal covert channels on multi-core platforms,''
  in \emph{Proc. USENIX Sec. Symp.}, 2015, pp. 865--880.

\bibitem{ITRS15}
\BIBentryALTinterwordspacing
(2016) International technology roadmap for semiconductor 2.0. {ITRS}.
  [Online]:
  \url{http://www.semiconductors.org/main/2015_international_technology_roadmap_for_semiconductors_itrs/}
\BIBentrySTDinterwordspacing

\bibitem{kim12_3dmaps}
D.~H. Kim \emph{et~al.}, ``{3D-MAPS}: {3D} massively parallel processor with
  stacked memory,'' in \emph{Proc. Int. Sol.-St. Circ. Conf.}, 2012, pp.
  188--190.

\bibitem{iyer16}
S.~S. Iyer, ``Heterogeneous integration for performance and scaling,''
  \emph{Trans. Compon., Pack., Manuf. Tech.}, vol.~6, no.~7, pp. 973--982,
  2016.

\bibitem{beneventi16}
F.~Beneventi \emph{et~al.}, ``Thermal analysis and interpolation techniques for
  a logic + {WideIO} stacked {DRAM} test chip,'' \emph{Trans. Comp.-Aided Des.
  Integ. Circ. Sys.}, vol.~35, no.~4, pp. 623--636, 2016.

\bibitem{knechtel16_Challenges_ISPD}
J.~Knechtel and J.~Lienig, ``Physical design automation for {3D} chip stacks --
  challenges and solutions,'' in \emph{Proc. Int. Symp. Phys. Des.}, 2016, pp.
  3--10.

\bibitem{budhathoki16}
P.~Budhathoki \emph{et~al.}, ``Integrating {3D} floorplanning and optimization
  of thermal through-silicon vias,'' in \emph{{3D} Stacked Chips -- From
  Emerging Processes to Heterogeneous Systems}, I.~A.~M. Elfadel and
  G.~Fettweis, Eds.\hskip 1em plus 0.5em minus 0.4em\relax Springer, 2016.

\bibitem{samal16}
S.~K. Samal \emph{et~al.}, ``Adaptive regression-based thermal modeling and
  optimization for monolithic {3-D} {ICs},'' \emph{Trans. Comp.-Aided Des.
  Integ. Circ. Sys.}, vol.~35, no.~10, pp. 1707--1720, 2016.

\bibitem{zhu08}
C.~Zhu \emph{et~al.}, ``Three-dimensional chip-multiprocessor run-time thermal
  management,'' \emph{Trans. Comp.-Aided Des. Integ. Circ. Sys.}, vol.~27,
  no.~8, pp. 1479--1492, 2008.

\bibitem{fu17}
Y.~Fu \emph{et~al.}, ``Kalman predictor-based proactive dynamic thermal
  management for {3D} {NoC} systems with noisy thermal sensors,'' \emph{Trans.
  Comp.-Aided Des. Integ. Circ. Sys.}, vol.~PP, no.~99, pp. 1--1, 2017.

\bibitem{xie16}
Y.~Xie \emph{et~al.}, ``Security and vulnerability implications of {3D}
  {ICs},'' \emph{Trans. Multi-Scale Comp. Sys.}, vol.~2, no.~2, pp. 108--122,
  2016.

\bibitem{bao15}
C.~Bao and A.~Srivastava, ``{3D} integration: New opportunities in defense
  against cache-timing side-channel attacks,'' in \emph{Proc. Int. Conf. Comp.
  Des.}, 2015, pp. 273--280.

\bibitem{valamehr10}
J.~Valamehr \emph{et~al.}, ``Hardware assistance for trustworthy systems
  through {3-D} integration,'' in \emph{Proc. Ann. Comp. Sec. App. Conf.},
  2010, pp. 199--210.

\bibitem{gu16_ICCD}
P.~Gu \emph{et~al.}, ``Thermal-aware {3D} design for side-channel information
  leakage,'' in \emph{Proc. Int. Conf. Comp. Des.}, 2016, pp. 520--527.

\bibitem{paek13}
S.~Paek \emph{et~al.}, ``{PowerField}: A probabilistic approach for
  temperature-to-power conversion based on {Markov} random field theory,''
  \emph{Trans. Comp.-Aided Des. Integ. Circ. Sys.}, vol.~32, no.~10, pp.
  1509--1519, 2013.

\bibitem{Lim13}
S.~K. Lim, \emph{Design for High Performance, Low Power, and Reliable {3D}
  Integrated Circuits}.\hskip 1em plus 0.5em minus 0.4em\relax Springer, 2013.

\bibitem{knechtel_CorblivarCode}
\BIBentryALTinterwordspacing
J.~Knechtel. (2017) Corblivar floorplanning suite and benchmarks. [Online]:
  \url{https://github.com/IFTE-EDA/Corblivar}
\BIBentrySTDinterwordspacing

\bibitem{zhang15}
R.~Zhang \emph{et~al.}, ``{HotSpot} 6.0: Validation, acceleration and
  extension,'' University of Virginia, Tech. Rep., 2015.

\bibitem{demme12}
J.~Demme \emph{et~al.}, ``Side-channel vulnerability factor: A metric for
  measuring information leakage,'' \emph{SIGARCH Comp. Arch. News}, vol.~40,
  no.~3, pp. 106--117, 2012.

\bibitem{claramunt05}
C.~Claramunt, ``A spatial form of diversity,'' in \emph{Proc. Int. Conf.
  Spatial Inf. Theory}, 2005, pp. 218--231.

\bibitem{knechtel15_Corblivar_TCAD}
J.~Knechtel \emph{et~al.}, ``Planning massive interconnects in {3-D} chips,''
  \emph{Trans. Comp.-Aided Des. Integ. Circ. Sys.}, vol.~34, no.~11, pp.
  1808--1821, 2015.

\bibitem{ma11_MSV}
Q.~Ma \emph{et~al.}, ``{MSV}-driven floorplanning,'' \emph{Trans. Comp.-Aided
  Des. Integ. Circ. Sys.}, vol.~30, no.~8, pp. 1152--1162, 2011.

\bibitem{lin10}
H.-L. Lin, ``A multiple power domain floorplanning in {3D IC},'' Master's
  thesis, National Tsing Hua University, Taiwan, 2010.

\bibitem{lee14}
B.~Lee \emph{et~al.}, ``Voltage islanding technique for concurrent power and
  temperature optimization in {3D-stacked ICs},'' in \emph{Proc. Int. Conf.
  Circ. Sys. Comp. Comm.}, 2014, pp. 267--269.

\end{thebibliography}

\end{document}